\documentclass[showpacs,showkeys,reprint,groupedaddress]{revtex4}
\usepackage{amsmath}
\usepackage{graphicx}
\usepackage{amssymb}
\usepackage{epstopdf}
\usepackage{bm}

\providecommand{\U}[1]{\protect\rule{.1in}{.1in}}
\begin{document}

\title{Optimizing Nanoparticle Designs to Reach \\ Ideal Light Absorption}
\author{Victor Grigoriev, Nicolas Bonod, Jerome Wenger, Brian Stout}
\affiliation{Aix-Marseille Universit\'{e}, Institut Fresnel, Facult\'{e} des Sciences de Saint J\'{e}r\^{o}me, 
13397 Marseille, France }
\date{\today }

\begin{abstract}
Ideal absorption describes a particular means of optimizing light-matter interactions with a host of potential
applications. This work presents new analytic formulas and describes semi-analytical methods  
for the design of electric or magnetic ideal absorption in nanoparticles. These formulas indicate that ideal absorption 
is attainable in homogeneous spheres with known materials at specific sizes and frequencies. They also provide 
a means of designing core-shell particles to produce ideal absorption at virtually any frequency in the visible and 
near infrared range.
\end{abstract}

%\ocis{260.1960,260.2110,300.1030}  % 260.1960 Diffraction theory, 260.2110   Electromagnetic optics
%260.2110   Electromagnetic optics 
%300.1030   Absorption
%insert suggested PACS numbers in braces on next line

\maketitle

\affiliation{Institut Fresnel, CNRS, Aix-Marseille Universit\'{e}, Ecole
Centrale Marseille \\ Campus de Saint-J\'{e}r\^{o}me, 13013 Marseille,
France}

%insert suggested keywords - APS authors don't need to do this \keywords{}


\section{Introduction}
Subwavelength sized particles that feature electromagnetic resonances are ideal tools to concentrate light at the nanometer scale. They play an important role in a wide range applications\cite{Lal07,Sch10} including receiving elements in optical antennas,\cite{Nov11} solar panel transducers\cite{Atw10} and sensing\cite{May11}. In this context, the design of nanoparticles that feature the ability to optimally convert light energy in at least one incident field mode into another form like photoluminescence or heat is of crucial importance.\cite{Ha11} Such particles will be defined in this study as Ideal Absorption (IA) particles. Ideal absorption has also been called coherent perfect absorption\cite{Stone:10,Noh:12,Noh:13} since 
it results in a mode of the \emph{total field} taking the form of a purely incoming being \emph{completely} 
absorbed by the particle. In this work, we prefer to avoid this terminology since absorption is only \lq perfect\rq\ for 
an ideal $4\pi$ illumination specifically adapted to the corresponding IA mode.\cite{Sentenac:13} 

%This work derives convenient formulas enabling the design of IA in spherically symmetric sub-wavelength particles
%whose symmetry allows IA to simultaneously occur in all the degenerate orientation eigenstates, $m$, associated with %each order angular momentum number, $n$.

In this work, ideal absorption is first introduced by demonstrating that it corresponds 
to a fundamental upper bound in an absorption cross section channel, 
($\sigma_{\mathrm{a}}=3\lambda^{2}/8\pi$ for a \emph{dipole} resonator). A `point-like' model is then shown 
to provide an analytical formula for predicting an electric dipole IA, but since this model is inaccurate for all but 
the smallest particles, significantly improved approximate formulas are derived. We proceed to show that the ensemble
of IA solutions can be determined using the so called Weierstrass factorization of 
the analytical scattering response of spherical particles.\cite{Vict13} Other considerations, like multi-mode absorption
and IA bandwidth are discussed using illustrative calculations. We demonstrate that our formulas predict IA in
homogeneous spheres with realizable materials at specific frequencies, and sizes. This work finishes with a derivation of a simple procedure for designing core-shell particles to exhibit IA at essentially any frequency in the visible and near visible frequency range.

\section{Scattering theory for Ideal Absorption}
Scattering in three dimensional electromagnetic problems can conveniently be
expressed in terms of incoming and outgoing spherical Vector Partial Waves (VPWs).\cite{Ts00}
The VPWs are solutions of homogeneous media Maxwell equations of either electric source type 
($e$ : $\bm{N}_{n,m}^{\left( \pm \right) }\left( k\mathbf{r}\right) $) or magnetic source type 
($h$ : $\bm{M}_{n,m}^{\left( \pm \right) }\left( k\mathbf{r}\right) $), with the $+(-)$ superscripts 
indicating that the functions satisfy outgoing (incoming) boundary conditions respectively. The subscript,
$n$, denotes the total angular momentum number, and $m=-n,...,n$, the angular momentum projection number.\cite{Ts00}
The spatial dependence of the VPWs is scaled by the in-medium wavenumber, 
$k=\sqrt{\varepsilon_b \mu_b}\omega/c= N_{b}\omega/c = 2\pi/\lambda$, where $N_{b}$ is the
refraction index of the homogeneous background material, and $\lambda$, the in-medium wavelength. 

Theoretical treatments of electromagnetic scattering from a homogeneous spherically symmetric particle
go under the names of Lorenz-Mie-Debye theory, but the physical content is best viewed in the
formalism of S or T matrices, originally developed for quantum mechanical scattering 
theory.\cite{Hu57,Ts00,Ne66,St08,Bohr83,Niem:03} The S-matrix seems best adapted to the discussion of IA, and 
is privileged from here on. 

The \emph{total field} in a homogeneous region surrounding a particle can always be developed on the basis set of 
the incoming and outgoing VPWs, with $a_{n,m}^{(e,\pm) }$ and $a_{n,m}^{(h,\pm)}$ respectively
denoting the the electric and magnetic mode VPWs field coefficients. Since S-matrix of 
a spherically symmetric system is automatically diagonal in the VPW basis, its elements express the linear relationship
between the outgoing and incoming field coefficients of the total field for both electric and magnetic source fields respectively, 
\textit{i.e.} $S_{n}^{\left( e\right)}=a_{n,m}^{(e,+) }/a_{n,m}^{(e,-) }$, and 
$S_{n}^{\left( h\right)}=a_{n,m}^{(h,+) }/a_{n,m}^{(h,-) }$.

Algebraic manipulations involving the T-matrix\cite{St08,Ts00}, and 
the definition of the S-matrix\cite{Ts00,Hu57}, 
$\overline{\overline{S}}\equiv \overline{\overline{I}}+2\overline{\overline{T}}$, 
provide convenient expressions for the S-matrix coefficients:
\begin{subequations}
\begin{align}
S_{n}^{\left(  e\right)} &  =-\frac{h^{(-)}_{n}\left(  kR\right)}
{h^{(+)}_{n}\left(kR\right)} \cdot \frac{ \overline{\varepsilon}_{s}
\varphi_{n}^{\left(-\right) }\left(  kR\right) -\varphi_{n}\left(  k_{s}R\right)}
{\overline{\varepsilon}_{s}\varphi_{n}^{\left(  +\right)  }
\left(kR\right) -\varphi_{n}\left( k_{s}R\right)}  \label{Se}\\
S_{n}^{\left(  h\right)} &  =-\frac{h^{(-)}_{n}\left(  kR\right)}
{h^{(+)}_{n}\left(kR\right)} \cdot \frac{ \overline{\mu}_{s}
\varphi_{n}^{\left(-\right) }\left(  kR\right) -\varphi_{n}\left(  k_{s}R\right)}
{\overline{\mu}_{s}\varphi_{n}^{\left(  +\right)  }
\left(kR\right) -\varphi_{n}\left( k_{s}R\right)}\ , \label{Sh} 
\end{align}
\label{MieS}
\end{subequations}

\noindent where, $\overline{\varepsilon}_{s} \equiv \varepsilon_{s}/\varepsilon_{b}$, and 
$\overline{\mu}_{s} \equiv \mu_{s}/\mu_{b}$ are respectively the permittivity
and permeability contrasts between the sphere and the host medium, and $k_{s}$ is the 
wavenumber inside the sphere. The $h^{(+)}_{n}(x)$ and $h^{(-)}_{n}(x)$ functions
respectively denote the outgoing and incoming spherical Hankel functions. 
The expressions in Eq.(1) also employ the functions: %\ref{MieS}
\begin{equation}
\varphi_{n}^{(\pm)}(x)  \equiv \frac{\left[xh_{n}^{\left(\pm\right)}(x)\right]^{\prime}}
{h_{n}^{\left(\pm\right) }(x)} \ , \qquad \varphi_{n} (x)  \equiv 
\frac{\left[xj_{n}(x)\right]^{\prime}}{j_{n}(x)} \ ,
\end{equation}
where $j_{n}\left( x\right) $ are the $n^{th}$ order spherical Bessel functions. 

Flux conservation imposes an upper bound on the amplitude of the $S$ matrix elements, 
$\left\vert S_{n}^{\left(e,h\right) }\right\vert \leq1$,
where the upper limit, $\left\vert S_{n}^{\left(e,h\right)}\right\vert =1$ is satisfied by any lossless 
scatterer since this condition results directly in 
$\left\vert a_{n,m}^{(+)}\right\vert =\left\vert a_{n,m}^{(-)}\right\vert $.\cite{Hu57}
The S-matrix of a  lossless scatterer is characterized by zeros (absorbing modes) in the upper-half  complex frequency
plane, and poles (emitting modes) in the lower-half plane frequency plane.\cite{Vict13} Adding absorption to the
particle causes the absorbing modes to descend towards (and finally into) the lower complex plane with IA occurring 
at those values of $\overline{\varepsilon}_{s}$ for which a zero of the S-matrix lies on the real frequency axis, \textit{i.e.} there exists an $n$ such that:
\begin{subequations}
\begin{align}
S_{n}^{\left( e\right) }&=0 \Leftrightarrow \frac{\varphi_{n}\left(  k_{s}R\right)}
{\overline{\varepsilon}_{s}}  = \varphi_{n}^{\left(-\right)} \left(  kR\right) \ , \   {\rm or} \ , \label{eIA}\\
S_{n}^{\left( h\right) }&=0 \Leftrightarrow \frac{\varphi_{n}\left(  k_{s}R\right)}
{\overline{\mu}_{s}} =\varphi_{n}^{\left(-\right)} \left( kR\right)   \ , \label{hIA}
\end{align} \label{abscond}
\end{subequations}
when $\Im\{\omega \}=0$ (these conditions being found by inspection of Eq.(\ref{Se}) and Eq.(\ref{Sh}).

Scattering and absorption cross sections of a spherically symmetric scatterer  
are the sum of the contributions from all multi-pole orders, $n$:
\begin{equation}
\sigma _{\mathrm{s}}=\sum_{n=1}^{\infty}\left( \sigma _{n,\mathrm{s}}^{\left( e\right)}
+\sigma_{n,\mathrm{s}}^{\left( h\right)}\right) \ ,\quad \sigma_{\mathrm{a}}
=\sum_{n=1}^{\infty}\left( \sigma_{n,\mathrm{a}}^{\left( e\right)}+
\sigma_{n,\mathrm{a}}^{\left( h\right) }\right) \ .
\end{equation}
The multi-pole contributions to the extinction, scattering, and absorption cross sections\cite{Ts00,Hu57,Bohr83} 
can be respectively expressed in terms of the S-matrix as:
\begin{subequations}
\begin{align}
\sigma_{n,\mathrm{a}}^{\left( e,h\right) }& =\frac{\pi }{2k^{2}}
\left(2n+1\right) \left( 1-\left\vert S_{n}^{\left( e,h\right)}\right\vert^{2}\right) \label{abssect} \\
\sigma_{n,\mathrm{s}}^{\left( e,h\right) }& =\frac{\pi }{2k^{2}}
\left(2n+1\right) \left\vert S_{n}^{\left( e,h\right) }-1\right\vert ^{2}  \label{scatsect} \\
\sigma_{n,\mathrm{e}}^{\left( e,h\right) }& =\frac{\pi }{k^{2}}
\left(2n+1\right) \Re\left\{ 1-S_{n}^{\left( e,h\right) }\right\} \ , \label{extsect}
\end{align}
\label{crossect}
\end{subequations}
where  the $2n+1$ factors arise from the azimuthal mode degeneracy of the orbital modes.

One finds that an IA mode contributes equally to the absorption and scattering 
cross sections by inserting the IA criteria, $S_{n}=0$ into Eq.(\ref{abssect}) and Eq.(\ref{scatsect}) (in either the $(e)$ or $(h)$ 
channels), with values of:
\begin{equation}
\sigma_{n,\mathrm{s}} =
\sigma_{n,\mathrm{a}} =\frac{2n+1}{8\pi }\lambda^{2}\ ,  \label{crosslim}
\end{equation}
which is an upper bound for the absorption cross section of a mode, but a factor four times less
than the $S_{n}=-1$, \emph{unitary limit}\cite{Ne66} of the scattering or extinction
cross sections, obtainable from Eq.(\ref{scatsect}) and Eq.(\ref{extsect}). 
The equality of scattering and absorption cross sections in an IA channel follows from 
the fact that IA requires the field scattered by the particle to have perfectly
destructive interference with the outgoing wave components of the local excitation
field (leaving the $n^{th}$ order modes of the total field to be purely incoming). We also remark 
the many properties of IA in particles, like Eq.(\ref{crosslim}), find analogues in the ideal absorption of 
1D systems, where for example the maximum absorption coefficient, $A$, of a plane wave illuminating a single side 
of a symmetric thin film occurs when $A=R+T=0.5$.\cite{Bott:97,Sarychev:95}

Since the $\varphi_{n}^{\left(-\right)} \left(  kR\right)$ is a complex 
function of a real variable, and  $\varphi_{n} \left(  k_s R\right)$ is real for real valued $k_s$, the IA
solutions of Eq.(\ref{eIA}) or Eq.(\ref{hIA}) can occur only for complex values of $\overline{\varepsilon}_{s}$.
Furthermore, Eqs.(3) are transcendental equations, and each has % \ref{abscond} 
an infinite number of solutions. For the sake of clarity, further discussion in this work will be limited 
to the $n=1$ (\textit{i.e.} dipole) mode solutions of Eq.(3). % \ref{abscond} 

\begin{figure}[!htb]
\includegraphics[width=0.8\linewidth]{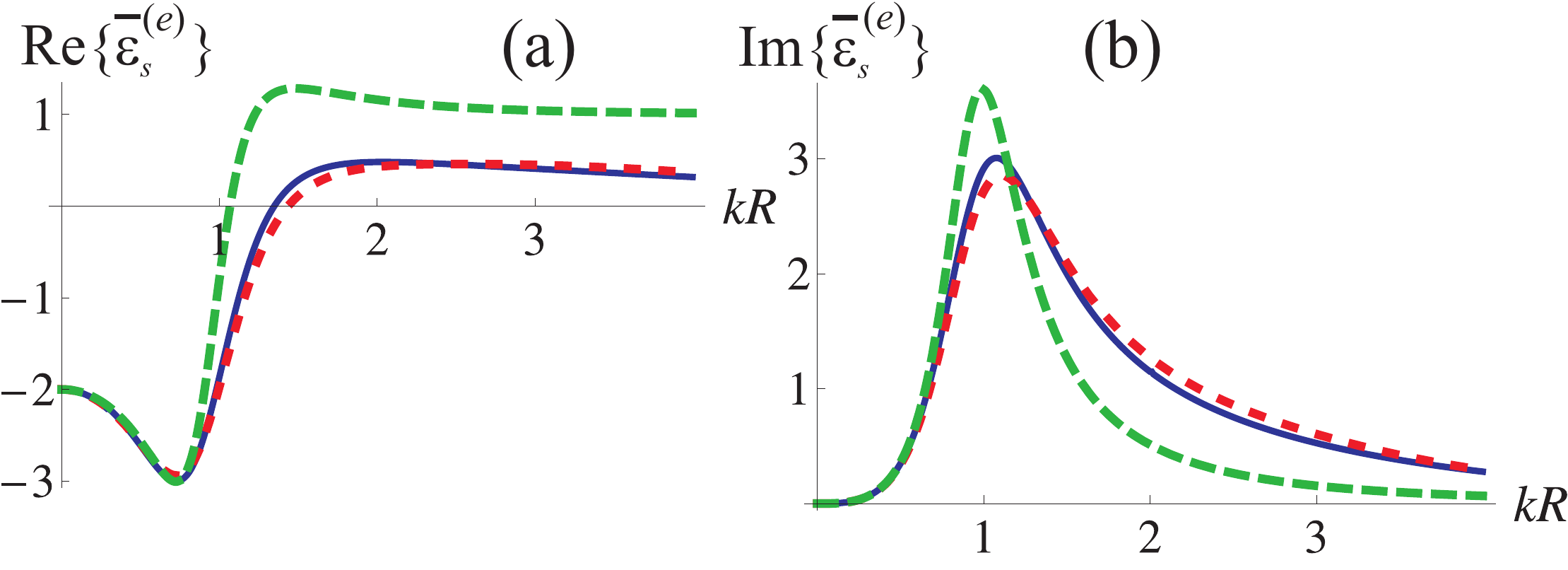}
\caption{\label{fig:IAapprox} (Color on-line) The real and imaginary parts of
$\overline{\varepsilon}_{s}$ yielding the lowest IA solution are plotted as a function of $kR$ 
(blue (solid) curves in (a) and (b)).  The point-like model prediction of Eq.(\ref{greenpoint}), is plotted in 
green (long dashed), while the improved approximation of Eq.(\ref{approxsol}) is plotted in red (short dashed).}
\end{figure}

The real and imaginary parts of the solutions of Eq.(\ref{eIA})  
with the lowest values of $\Re\{\overline{\varepsilon}_{s}\}$ are plotted as the blue(solid) line in 
Fig.(\ref{fig:IAapprox}) for particle diameters ranging from zero up to a little more than a wavelength. 
These results were computed 
by taking advantage of the newly developed Weierstrass factorization for resonant photonic 
structures.\cite{Vict13} One readily remarks that this solution tends to the well-known quasi-static
dipolar plasmon resonance at $\overline{\varepsilon}_{s}=-2$ when $kR\to 0$. When this is the only
IA mode of interest, one can try to describe IA with a `point-like' model aimed at providing
approximate descriptions of the lowest $\Re\{\overline{\varepsilon}_{s}\}$ electric dipole 
resonance\cite{lag98,pe97b}.

\section{Point-like models and analytic extensions}

Point-like models have generally been derived from first principles
using Green's function theory, and the most rigorous derivations predict 
a point-like electric dipole polarizability, $\alpha$, that can be 
written:\cite{pe97b,ColasdesFrancs:07,ColasdesFrancs:08}
\begin{equation}
\left[ {\alpha}^{(e)} \right]^{-1}=\alpha_{0}^{-1}-\frac{k^{2}}{6\pi R}-i\frac{k^{3}}{6\pi}\ , \quad \alpha_{0}\equiv 4\pi R^3 
\frac{\overline{\varepsilon}_{s}-1 }{\overline{\varepsilon}_{s}+2} \ , \label{oldpoint}
\end{equation}
where $\alpha_{0}$ is the quasi-static polarizability of a spherical particle. The $-ik^3/6\pi$ term in
Eq.(\ref{oldpoint}) has been extensively studied due to its importance in energy conservation, while the
$-k^2/ 6\pi R$ term has received considerably less attention even though it is responsible for the 
well-known `red-shift' of the localized plasmon resonance with increasing particle size. 

For a spherically symmetric scatterer, one can directly connect the isotropic polarizability in terms of the 
electric dipole S-matrix element, $S_{1}^{\left(  e\right)}$, via the 
relation:\cite{Stout11}
\begin{equation}
\alpha^{(e)}= \frac{3\pi i}{k^{3}}\left( 1 - S_{1}^{\left(  e\right)} \right) \ , \label{Salph}
\end{equation}
leading to an IA polarizability of $\alpha_{\rm IA}= {3\pi i}/{k^{3}}$. An
advantage of the point-like model is that one can algebraically solve Eq.(\ref{oldpoint}) for the value of
$\overline{\varepsilon}_{\rm IA}^{(e)}$ producing an IA polarizability as a function of the size parameter, 
$\rho\equiv kR$:
\begin{align}
\overline{\varepsilon}_{\rm IA}^{(e)} (\rho) &  =-\frac{2+\frac{2}{3}\rho^{2}
\left(1-i\rho\right)}{1-\frac{2}{3} \rho^{2}
\left(1-i\rho\right)} \ . \label{greenpoint}
\end{align}
The real and imaginary parts of this expression for $\overline{\varepsilon}_{s}$ are plotted in 
Fig.(\ref{fig:IAapprox}a)  and Fig.(\ref{fig:IAapprox}b) (green curves), and although 
$\overline{\varepsilon}_{s}$ tends towards the numerically obtained result (blue curve) in the
$kR\to 0$ limit, it differs significantly from the exact result for larger particle sizes. 

One can determine much more reliable predictions of the IA conditions by analyzing the pole 
structure of the special functions. One first remarks that the $\varphi_{n}^{\left(\pm \right)}$ 
functions appearing in Eq.(\ref{MieS}) have finite meromorphic expressions, which for the dipole example take the form:
\begin{equation}
h_{1}^{\left(  \pm\right)  }\left(  z\right)  =-e^{\pm iz}\frac{z\pm i}{z^{2}} \  ; \
\varphi_{1}^{\left(\pm\right)}\left( z\right)  =\pm iz-\frac{1}{1\mp iz} \ .
\end{equation}
The $\varphi_{n}$ functions on the other hand have an infinite number of poles located along the 
real axis, and can be expressed in meromorphic form as:
\begin{equation}
\varphi_{n}\left(  z\right)  = n+1 + \sum_{\alpha=1}^{\infty}\left( \frac{2z^{2}}{z^{2}
-a_{n,\alpha}^{2}}\right) \ , \label{phi1}
\end{equation}
where the constants, $a_{n,\alpha}$, are the zeros of the spherical Bessel function, $j_{n}(x)$,
and are tabulated.\cite{Wat80}

One can obtain analytic dipole approximations to Fig.(\ref{MieS}) by replacing $\varphi_{1}$ of Eq.(\ref{phi1}) 
with an approximate meromorphic function having the same two lowest  poles, $a$ and zeros, $b$:
\begin{equation}
\varphi_{1}\left(  z\right)\approx 2\frac{1-(z/b)^{2}}{1-(z/a)^{2}} \ ,
\end{equation}
where $a=±1.4303\pi$ and $b=±0.87335\pi$. Adopting this substitution transforms $S_{1}^{(e,h)}=0$ into quadratic equations in terms of $\overline{\varepsilon}_{s}^{(e,h)}$ whose relevant solutions are:
\begin{subequations}
\begin{align}
\overline{\varepsilon}_{\rm IA}^{(e)}\left(\rho\right)&=\frac{1}{2}\frac{b^{2}}{ \rho^{2}}
\left[1+\frac{\rho^2}{a^2}  \frac{2 \left(i \rho+1\right)}{ \left(\rho ^2 -i\rho -1\right)}\right. \notag \\ 
& \left. -\sqrt{\left(1+\frac{\rho^2}{a^2}
 \frac{2 \left(i \rho+1\right)}{\left(\rho^2 -i\rho -1\right)}\right)^2-\frac{\rho^2}{b^2}  
\frac{8 \left(i \rho+1\right)}{\left(\rho^2 -i\rho -1\right)}}\right] \label{eIAmod} \\
\overline{\varepsilon}_{\rm IA}^{(h)}\left(\rho\right)&=\frac{3+3 i \rho - \rho^2}{\rho^2 
\left(\frac{2 + 2 i \rho}{b^2}+\frac{1+i \rho-\rho^2}{a^2}\right)} \ . \label{hIAmod}
\end{align}
\label{approxsol}
\end{subequations}

The electric dipole solution of Eq.(\ref{eIAmod}) is plotted in red in Fig.(\ref{fig:IAapprox}), while the plot of 
Eq.(\ref{hIAmod}) is indistinguishable on the scale of the numerically exact solution plotted in blue in 
Fig.(\ref{fig:IAcond}c) and Fig.(\ref{fig:IAcond}d).

\section{Exact solutions}

Although the analytic expressions of Eq.(13) provide accurate approximations for the %\ref{approxsol}
lowest IA modes, one may nevertheless desire higher precision or the predictions of additional IA
solutions, either of which will require numerical solutions of Eq.(\ref{eIA}) or Eq.(\ref{hIA}).  Finding %\ref{abscond}
these solutions is quite difficult when using commonly employed techniques involving
Cauchy integrals in the complex plane or conjugate gradient methods; but they can be readily solved using
techniques based on Weierstrass factorization\cite{Vict13} and which exploit the meromorphic expansion of 
Eq.(\ref{phi1}).

\begin{figure}[!htb]
\includegraphics[width=0.8\linewidth]{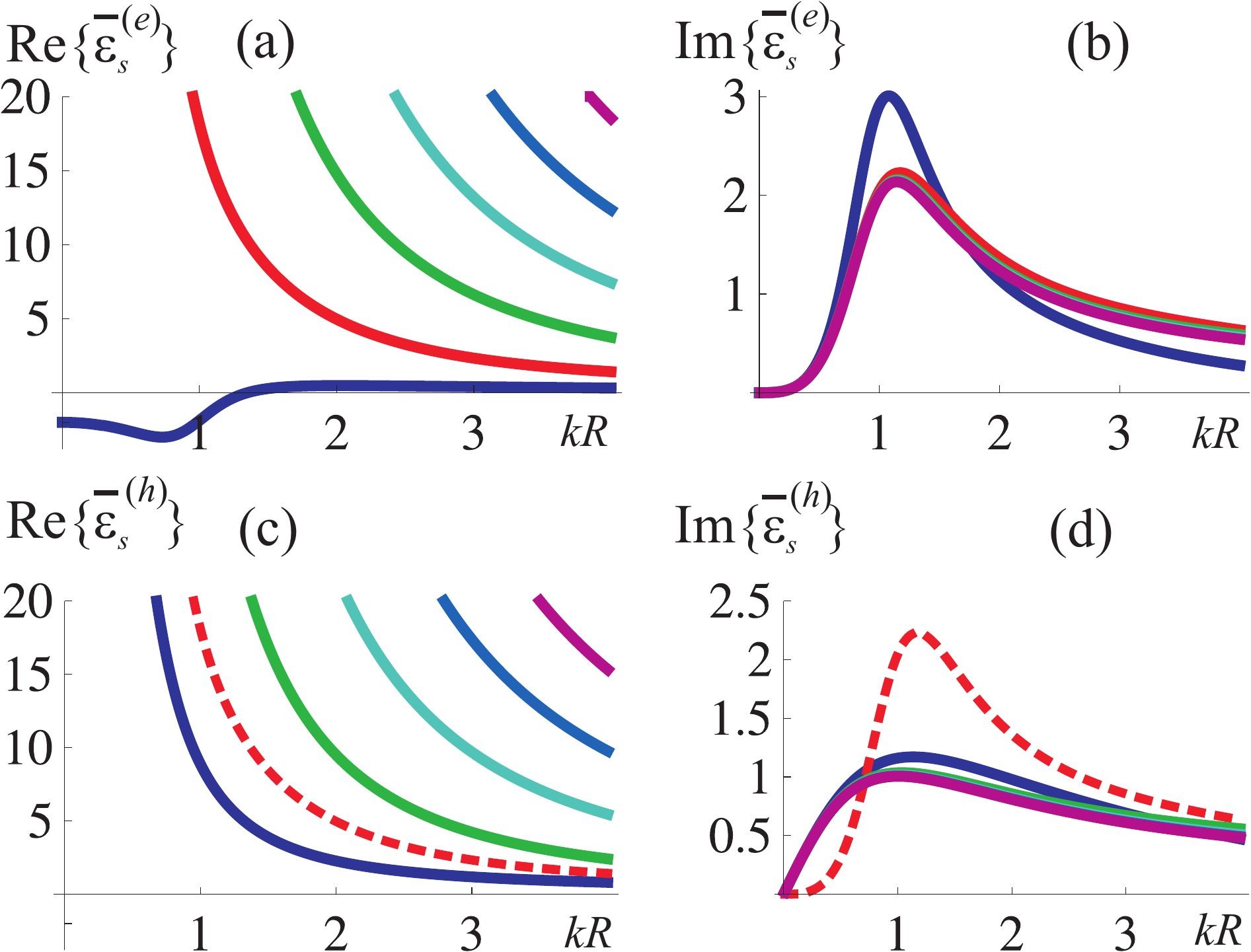}
\caption{\label{fig:IAcond} (Color on-line) The real, (a), and imaginary parts, (b), of  
the electric dipole IA conditions on $\overline{\varepsilon}_{s}$ are plotted as a function of 
$kR$.  Magnetic dipole IA values for $\overline{\varepsilon}_{s}$, are also plotted  as functions of $kR$
in  (c)-Re$\{\overline{\varepsilon}\}$ and (d)-Im$\{\overline{\varepsilon}\}$. The lowest electric dipole 
solution with $\Re \left\{ \overline{\varepsilon}_{s}\right\}>1$ is plotted as a dashed curve in 
(c) and (d) for the purpose of comparison.}
\end{figure}

The real and imaginary parts of the numerically obtained dipole solutions of Eq.(\ref{eIA}) and Eq.(\ref{hIA}) for
the $\overline{\varepsilon}_{s}$ are plotted in Fig.(\ref{fig:IAcond}) as a function of $kR$ for 
particles ranging in size from zero up to slightly larger than a wavelength ($kR=\pi$ corresponds to
particle  diameter=$\lambda$).

\begin{figure}[!htb]
\includegraphics[width=0.8\linewidth]{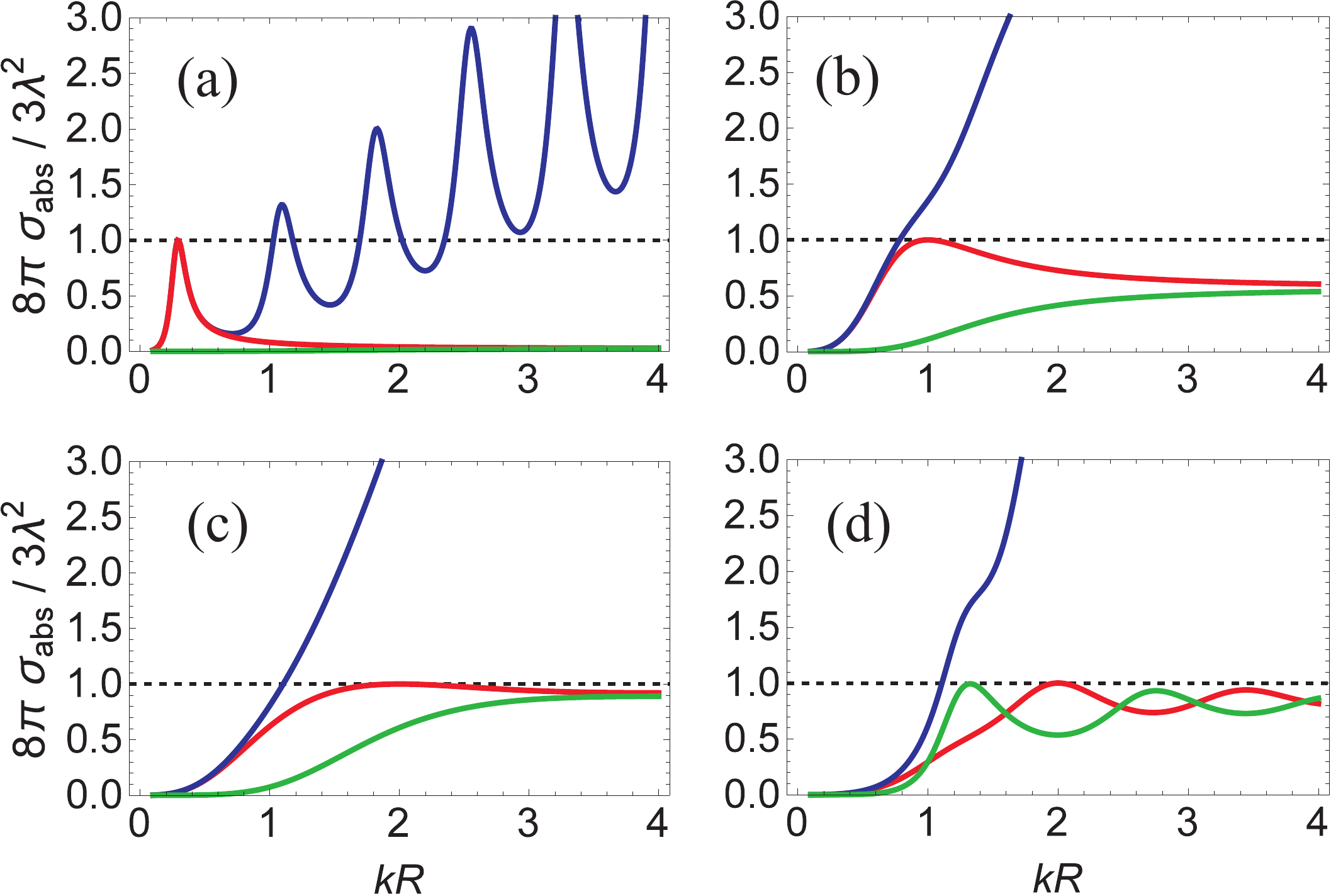}
\caption{\label{fig:sigabs} (Color on-line) Dimensionless absorption cross sections, 
$8\pi \sigma_{\rm abs}/ (3\lambda^{2})$,
plotted as a function of $kR$ for IA designed to occur at: (a) $kR=0.3$, 
$\overline{\varepsilon}_{s}=-2.21 + i 0.0611$; (b) $kR=1$, $\overline{\varepsilon}_{s} = -1.87 + i 2.91$; 
(c) $kR=2$, $\overline{\varepsilon}_{s}=0.4809 + i 1.148$; and (d) $kR=2$, 
$\overline{\varepsilon}_{s}=4.968 + i 1.361$. Total absorption cross sections are given by blue curves,
electric dipole absorption by red curves, and magnetic dipole absorption by green curves.}
\end{figure}

Even though the IA conditions can be found from the numerical solutions presented in Fig.(\ref{fig:IAcond}),  
absorber design will generally require taking into account the other properties of the absorber, like IA bandwidth
and absorption in other modes. We illustrate this point with some plots of the dimensionless absorption cross
section, $8\pi \sigma_{\rm abs}/(3\lambda^{2})$, for particles designed to
satisfy electric dipole IA at a few different sub-wavelength size parameters, notably $kR=0.3$ and $kR=1$ in 
Fig.(\ref{fig:sigabs}a) and Fig.(\ref{fig:sigabs}b). The total absorption cross sections are plotted in blue, the electric 
dipole absorption cross sections in red, and as a further piece of information, the
magnetic dipole absorption cross section is plotted in green.

One remarks from Fig.(\ref{fig:IAcond}) that IA can be satisfied for relatively modest values of 
$\Re \left\{\overline{\varepsilon}_{s} \right\}>1$ for size parameters roughly larger than unity
(\textit{i.e.} $D\gtrsim \lambda/3$). For IA designed to occur at $kR=2$, absorption cross
sections are plotted for both the $\Re\left\{\overline{\varepsilon}_{s}\right\}<1$ solution value of 
$0.4809 + i 1.148 $ in Fig.(\ref{fig:sigabs}c) and the lowest
$\Re\left\{\overline{\varepsilon}_{s}\right\}>1$ solution of  
$\overline{\varepsilon}_{s}=4.968 + i 1.361$ in Fig.(\ref{fig:sigabs}d).

For scatterers that are quite small with respect to the IA wavelength, like $kR=0.3$ in 
Fig.(\ref{fig:sigabs}a), one clearly observes peaks in the total absorption cross section associated with 
higher electric multi-pole modes, although only the electric dipole mode (red curve), satisfies the IA condition
per its design.
% (absorption in higher multi-pole orders can be stronger than that of an IA dipole mode on
%account of the $2n+1$ degeneracy factors) 
We remark that the magnetic dipole contributions (green curve) are insignificant at such sizes. For somewhat
larger sub-$\lambda$ particles, like the $kR=1$, the electric dipole IA solution in Fig.(\ref{fig:sigabs}b), 
one remarks that the electric dipole IA is accompanied by non-negligible absorption in higher order modes
including the magnetic dipole
contribution (in green). This behavior is accentuated for $kR=2$ with magnetic dipole contributions
coming close to the IA condition in both the $\Re\left\{\overline{\varepsilon}_{s}\right\}<1$ and
$\Re\left\{\overline{\varepsilon}_{s}\right\}>1$ designs. We remark however that even when IA 
occurs for both electric and magnetic dipole modes in the same particle, they cannot both be satisfied 
at the same frequency (for a homogeneous particle at least). We also underline the fact that for 
$\Re\{\varepsilon_{s}\}>1$ absorbers, it can be easier to obtain magnetic
dipole IA solutions than those of electric dipoles.  The ability for small particles to produce strong magnetic
dipole absorption at optical frequencies has also been remarked recently by other 
authors\cite{Asenjo-Garcia:12}.

\section{Ideal absorption with realistic materials}

Materials exhibiting plasmonic resonances, like silver and gold, are good candidates for achieving
ideal absorption at small particle sizes since they provide $\Re\left\{\overline{\varepsilon}_{s}\right\}<1$ and 
modest absorption over the visible and near visible frequencies. Even for these materials however, ideal absorption 
will only occur at certain frequencies and sizes, and generally requires exploiting at least one tunable parameter like 
the permittivity, $N_{b}$, of the background medium.

\begin{figure}[!htb]
\includegraphics[width=0.4\linewidth]{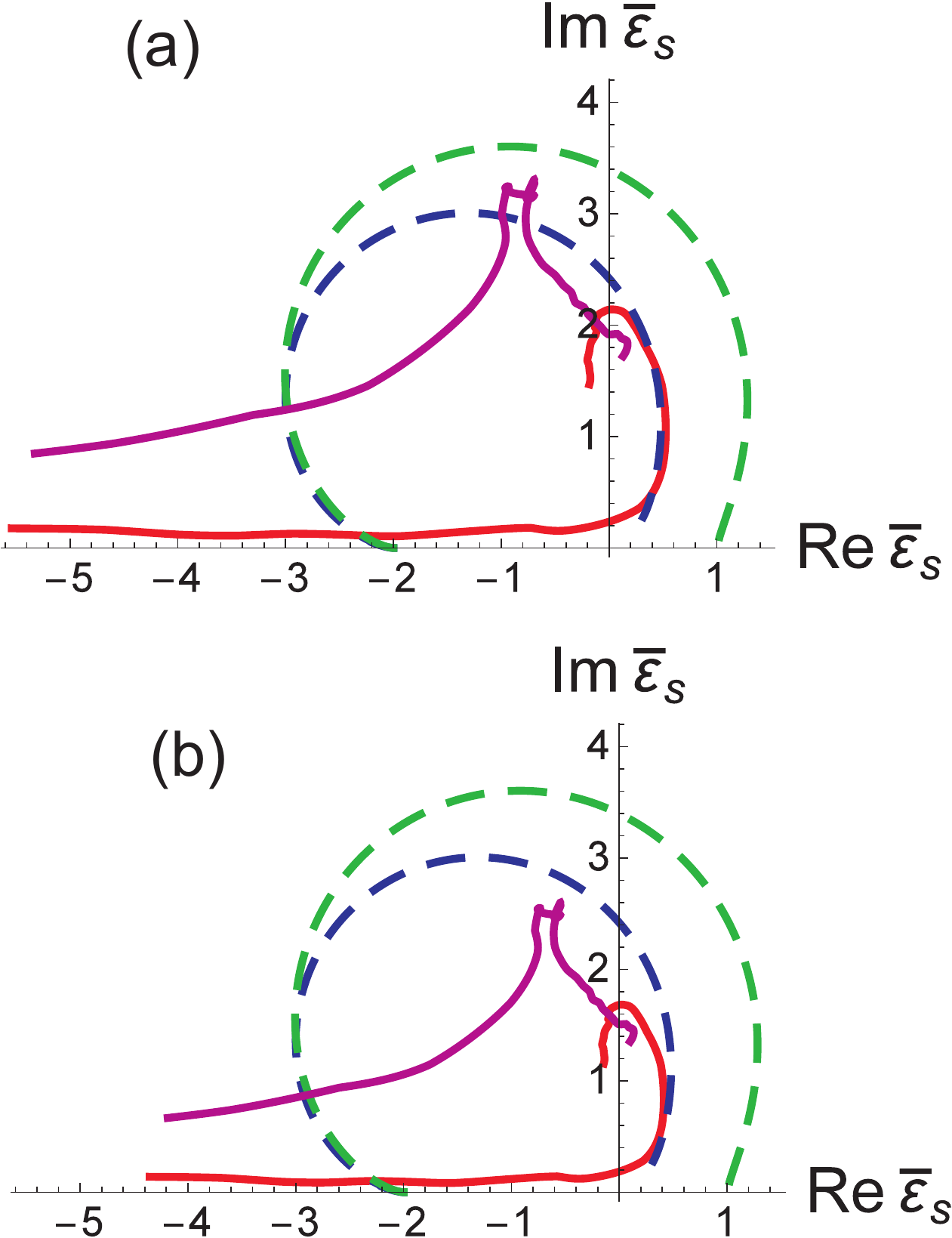}
\caption{\label{fig:silver_gold_IA} (Color on-line) Values of $\overline{\varepsilon}_{s}$ obtained by varying 
vacuum wavelength, for silver (red)  $\lambda_{v}\in\{188-500\}$nm and gold (magenta) 
$\lambda_{v}\in\{188-600\}$nm suspended in a (a) water background  medium, $N_b=1.33$, and (b) glass background  medium, $N_b=1.5$. The values of $\overline{\varepsilon}_{s}$ required for IA
in the lowest electric dipole mode are plotted in dashed blue for $kR$ varying from 0 to 4. Dipole model predictions of
Eq.(\ref{greenpoint}) are in plotted in dashed green. }
\end{figure}

We plot in Fig.(\ref{fig:silver_gold_IA}a), the path traced out by $\overline{\varepsilon}_{s}$ for gold and
silver in a water background, $N_{b}=1.33$, as the vacuum wavelength of light varies from the near ultra-violet to 
the mid-visible range (using interpolated Johnson \& Christy data\cite{Johnson_Christy:72}). The values of
$\overline{\varepsilon}_{s}$ required for producing the lowest electric dipole IA are also plotted in this figure as 
the size parameter, $kR$, varies from 0 to 4. The values predicted for IA by the point-like model are also plotted in 
this graph as $kR$ varies from 0 to infinity. IA is predicted to occur at the values of $kR$ and frequencies where the 
IA curves, dashed lines, intercept the experimentally determined dispersion relations, full lines. 

Although inspection of Fig.(\ref{fig:silver_gold_IA}a) indicates 3 possible IA solutions in a water background 
(cf. Fig.(\ref{tab:waterback})), we caution that only the lowest particle sizes correspond to resonances dominated by a dipolar 
response.  Dipole IA solutions at larger particles sizes will include significant absorption in higher order modes 
similar to that seen in Figs.(\ref{fig:sigabs}c-d).   

\begin{table}[!htb]
\begin{tabular}{|c|c|c|l|}
\hline
 $D_{\rm IA}$(nm) &$\lambda_{\rm v}^{\rm (IA)}\ ({\rm nm})$ & $kR_{\rm IA}$ & $\qquad \varepsilon_{\rm Ag}$ \\
\hline
34.04 & 393.76 & 0.361 &\ -4.094 + $i$0.199  \\
115.53 & 295.80 & 1.63 &\  \ 0.704 + $i$3.01  \\
291.4 & 321.0  &  3.13 &\ \ 0.700 + $i$0.856 \\
 \hline 
  $D_{\rm I}\ ({\rm nm})$ & $\lambda_{\rm v}^{\rm (IA)}\ ({\rm nm})$ & $kR_{\rm IA}$ & $\qquad \ \varepsilon_{\rm Au}$ \\
\hline
91.74 & 540.69 & 0.709 &\ -5.30 + $i$2.20  \\
121.59 & 451.81 &  1.12 &\  -1.76 + $i$5.27  \\
75.94 & 273.38 & 1.16 &\  -1.37 + $i$5.17 \\
\hline
\end{tabular}
\caption{ \label{tab:waterback} Particle diameters, $D_{\rm IA}$, and vacuum wavelengths, 
$\lambda_{\rm v}^{\rm (IA)}$, required to produce optical sinks of silver and gold 
in water, $N_b=1.33$.}
\end{table}

\begin{table}[!htb]

\begin{tabular}{|c|c|c|l|}
\hline
 $D_{\rm IA}\ ({\rm nm})$ & $\lambda_{\rm v}^{\rm (IA)}\ ({\rm nm})$ & $kR_{\rm IA}$ & $\qquad \ \varepsilon_{\rm Ag}$ \\
\hline
30.62 & 412.77 & 0.35 &\ -5.16 + $i$0.227 \\
 \hline
  $D_{\rm IA}\ ({\rm nm})$ & $\lambda_{\rm v}^{\rm (IA)}\ ({\rm nm})$ & $kR_{\rm IA}$ & $\qquad \ \varepsilon_{\rm Au}$ \\
  \hline
 76.23      & 559.68  &\ 0.64 &\ -6.57\  \ \,+ $i$1.95 \\
 \hline
\end{tabular}
\caption{\label{tab:glassback} Particle diameters, $D_{\rm IA}$, and  vacuum wavelengths, 
$\lambda_{\rm v}^{\rm (IA)}$, required to produce optical sinks of silver and gold 
in a $N_b$=1.5 background medium.}
\end{table}

Analogous curves are plotted in Fig.(\ref{fig:silver_gold_IA}b) for a higher background index material, 
$N_{b}=1.5$, like that typical of polymers and glass. At this higher index background medium, only one IA solution exists for both silver and gold. The IA values for gold and silver deduced from an analysis of the intercepts
between the IA conditions and the experimental permittivity functions of Fig.(\ref{fig:silver_gold_IA}) are given in
Table (\ref{tab:waterback}) and Table (\ref{tab:glassback}) for $N_b=1.33$ and $N_b=1.5$ respectively.

\begin{figure}[!htb]
\includegraphics[width=0.8\linewidth]{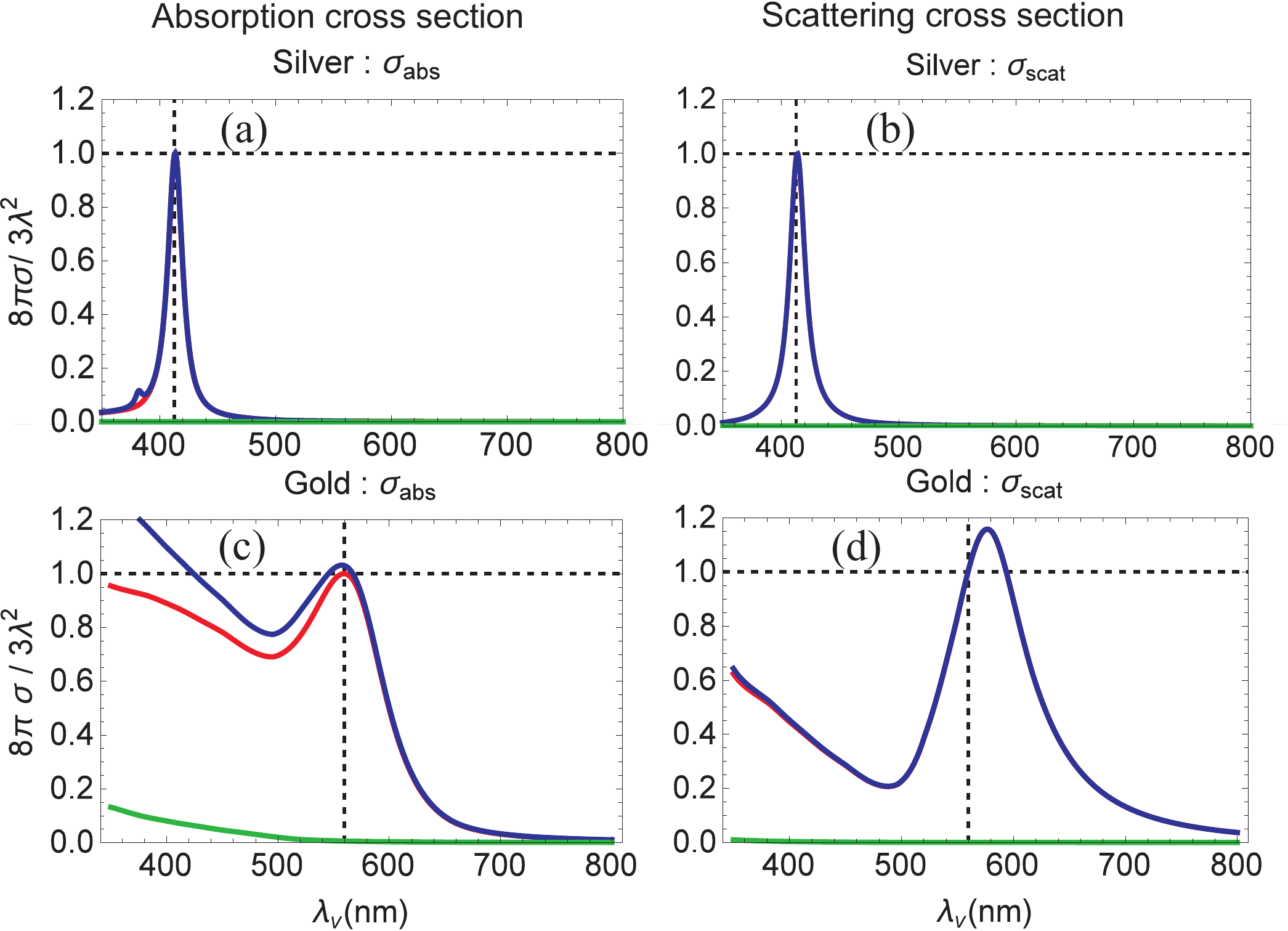}
\caption{\label{fig:silver_gold} (Color on-line) Cross sections (blue) of IA particles suspended in a 
$N_b=1.5$ background medium. Cross section for silver are plotted in : (a)-$\sigma_{\rm abs}$ 
and (b)-$\sigma_{\rm scat}$ and those for
gold in : (c)-$\sigma_{\rm abs}$ and (d)-$\sigma_{\rm scat}$ over the visible frequencies,
$\lambda_{v}\in\{350-800\}$nm.  Particle diameters are $D_{\rm IA}=30.6$nm for silver and $D_{\rm IA}=76.2$nm for gold, with their respective IA frequencies specified by vertical dashed lines. 
Dipole contributions to the cross sections are plotted in red for the electric dipole and green 
for the magnetic dipole.}
\end{figure}

Although IA is predicted for both silver and gold in transparent material media, one can expect 
significant differences in their IA behavior in view of the considerable differences in their respective
dispersions relations and predicted IA sizes.  
This is indeed the case, as illustrated in Fig.(\ref{fig:silver_gold}) where the scattering and absorption cross sections 
of the silver (a)-(b) and gold particles (c)-(d)  are plotted for frequencies in the visible range for particles whose IA 
diameters are taken from Table(\ref{tab:glassback}) ($D_{\rm IA}=30.6$nm ($\lambda_{v}=413$nm) for silver and 
$D_{\rm IA}=76.2$nm $\lambda_{v}=560$nm for gold). 

The frequencies at which IA is predicted are indicated by vertical dashed lines, and the IA cross sections of 
$8\pi\sigma_{\rm abs}/(3\lambda^2)=1$ are indicated by horizontal dashed lines in these figures. The total
absorption and scattering cross sections are drawn in blue, and electric dipole contributions are drawn in red, but these
curves are nearly indistinguishable on the scale of these graphs except for short wavelengths in gold. Magnetic dipole
contributions are plotted in green, but are negligible in all graphs excepting some mild magnetic dipole absorption
in gold at short wavelengths.

In Fig.(\ref{fig:silver_gold}), one remarks significant differences between the respective behaviors of silver and 
gold IA particles near their IA resonances. Notably, the electric quadrupole, magnetic dipole and higher order
corrections are essentially negligible near the silver IA resonance, while non-electric dipole orders contribute
significantly to absorption (but not scattering) at wavelengths below the IA resonance in gold. One should also 
remark that although the dipole contributions to the absorption cross sections are limited by their theoretical 
upper bound of $\sigma_{\rm abs}=3\lambda^2/(8\pi)$, the electric dipole scattering cross section of gold rises to 
values above $3\lambda^2/(8\pi)$, at frequencies below the IA resonance, but this is allowed by 
the general theoretical considerations of Eq.(\ref{crossect}) where one sees that the scattering cross section of a dipole mode 
is only required by unitarity to satisfy $\sigma_{\rm scat}\le 3\lambda^2/(2 \pi)$. Nevertheless, one can see in both 
Fig.(\ref{fig:silver_gold}b) and Fig.(\ref{fig:silver_gold}d) that $\sigma_{\rm scat}=3\lambda^2/(8 \pi)$ at the 
IA resonance as required by the general theoretical restraint of  Eq.(\ref{crosslim}).

\section{Ideal absorption for coated spheres}

Although we saw that changing the background material index allows some control over IA frequency with
homogeneous inclusions, applications are likely to want to design IA to occur at particular frequencies with a 
set of available materials. This will require additional adjustable geometric parameters and 
concentric coatings is one of the simplest ways to achieve this.

\begin{figure}[!htb]
\includegraphics[width=0.35\linewidth]{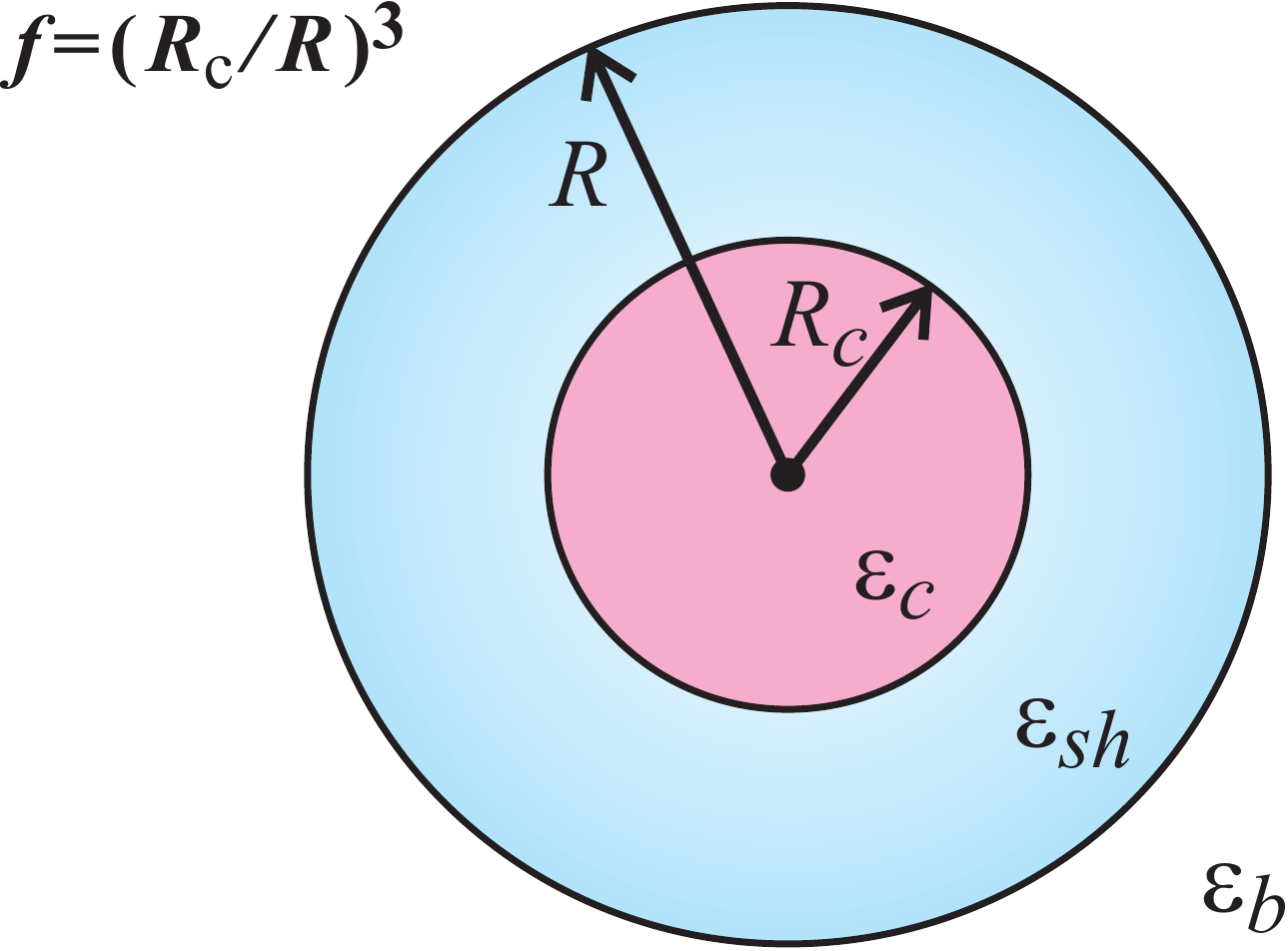}
\caption{\label{fig:coated_sch}(Color on-line) Schematic of a concentric coated sphere where the core, shell, 
and background permittivities are denoted respectively $\varepsilon_{c}$, $\varepsilon_{sh}$, and $\varepsilon_{b}$. The filling fraction of the core material is defined as $f\equiv\left(R_{c}/R\right)^{3}$.}
\end{figure}

If we consider inclusions consisting of a core and a single concentric coating, one has two adjustable parameters, 
the radius of the outer shell $R$, and its concentric spherical core, $R_{c}$ (cf. Fig.(\ref{fig:coated_sch}). 
It proves convenient to fix these two dimensioned parameters, $R$ and $R_{c}$, in terms of the dimensionless size
parameter, $kR$, and core material filling factor $f\equiv\left(R_{c}/R\right)^{3}$. Unlike the 
homogeneous sphere studied in the previous section, the analytic properties of the $S$ matrix cannot be directly
exploited to determine the parameters producing IA solutions, since no IA solutions are guaranteed to exist 
when varying the parameters $kR$ and $f$ of a coated sphere with fixed permittivities.

Given the complexity of the analytic coated sphere solutions, the possible non-existence of solutions at
finite frequency, and finally the fact that there are multiple adjustable parameters, it is advantageous to
develop compact approximate formulas that allow one to rapidly determine which parameters and
materials can provide viable designs for coated sphere IA. We derive now an approximation which
allows the design of core-shell electric dipole IA scattterers by taking advantage of the fact that one is generally
interested in designing sub-wavelength IA particles.

We saw that Eq.(\ref{eIAmod}) provides an accurate approximation for the required IA permittivity contrast,
$\overline{\varepsilon}_{s}$, over the entire range of the size parameter $\rho=kR$. The Maxwell-Garnett effective medium approach is derived as the effective permittivity of a concentric sub-wavelength core-shell particle
whose volume fraction matches that of the bulk material. The effective permittivity, 
$\overline{\varepsilon}_{\rm eff}$, of a core-shell system can thus be written as a function of filling fraction, $f$:\cite{Choy:99} 
\begin{equation}
\overline{\varepsilon}_{\rm eff}=\overline{\varepsilon}_{\rm sh} 
\left( 1+ 3 f \frac{ \overline{\varepsilon}_{c}-\overline{\varepsilon}_{sh}}
{\overline{\varepsilon}_{c}+2 \overline{\varepsilon}_{sh}-f (\overline{\varepsilon}_{c}-\overline{\varepsilon}_{sh})} \right) \ . \label{MaxG}
\end{equation}
where $\overline{\varepsilon}_{\rm sh}$ and $\overline{\varepsilon}_{\rm c}$ are the shell and core relative permittivities respectively.
This equation  can then be algebraically inverted to yields filling fraction as a function of 
$\overline{\varepsilon}_{\rm eff}$ {\it i.e.}:
\begin{equation}
f=\frac{\overline{\varepsilon}_{\rm eff}-\overline{\varepsilon}_{c}}
{\overline{\varepsilon}_{sh}-\overline{\varepsilon}_{c}}
\frac{\overline{\varepsilon}_{sh}+2 \overline{\varepsilon}_{c}}{\overline{\varepsilon}_{\rm eff}
+2 \overline{\varepsilon}_{c}} \ . \label{fcond}
\end{equation}

One can then solve for core-shell IA by replacing the ``effective'' index in Eq.(\ref{fcond}) by the $\overline{\varepsilon}^{(e)}$ of Eq.(\ref{eIAmod}) that specifies the permittivity required for the IA condition. 
Since $\overline{\varepsilon}_{\rm eff}$ and at least one of the materials is complex valued, the filling fraction,
$f$, found by Eq.(\ref{fcond}) is generally complex, but one can vary the size parameter, $\rho=k R$ to obtain 
a $\Im\{f\}=0$ solution graphically. Provided that the real part of the filling fraction is less than $1$ when 
$\Im\{f\}=0$, then one has found a valid IA solution.  

\begin{figure}[!htb]
\includegraphics[width=0.5\linewidth]{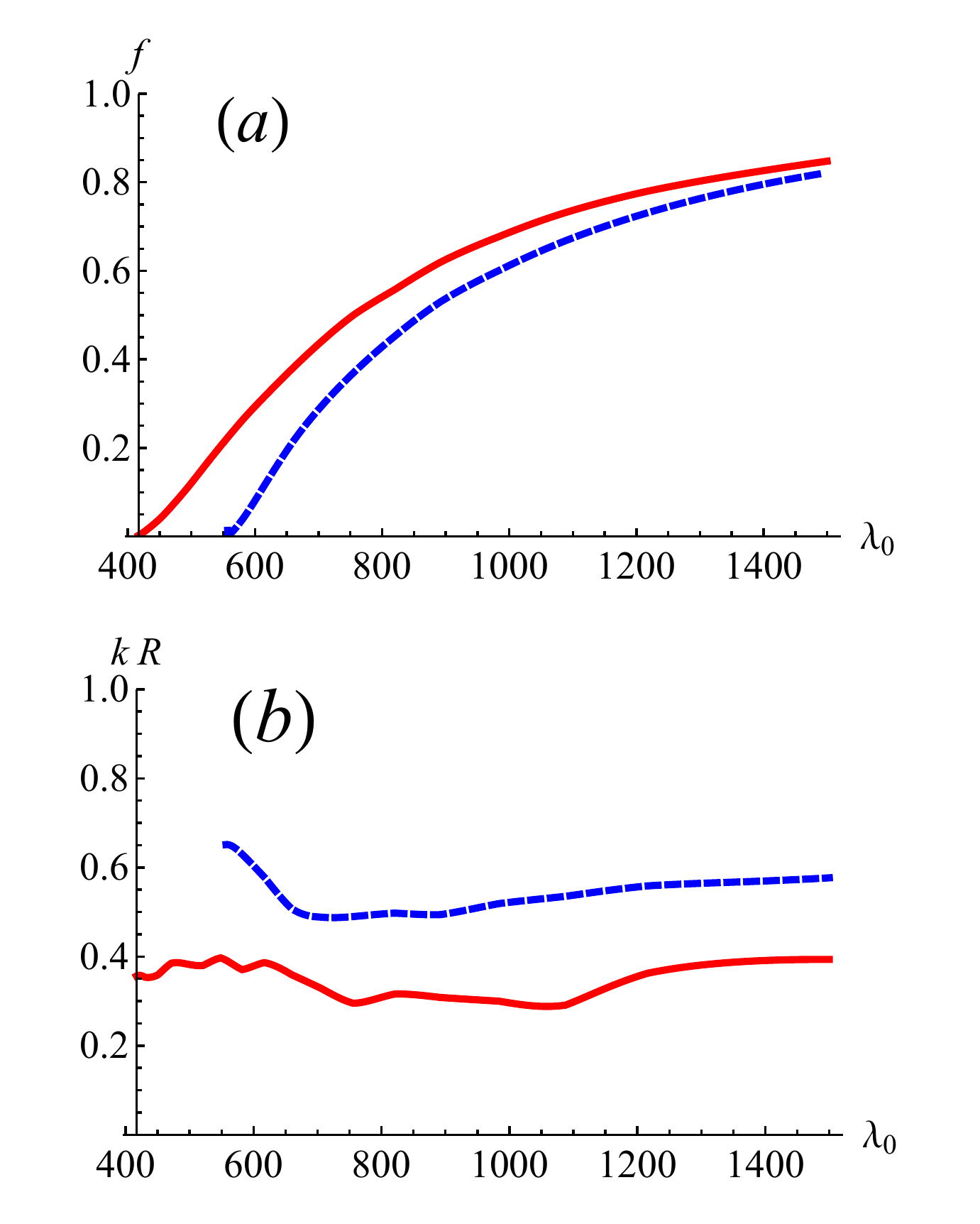}
\caption{\label{fig:core} (Color on-line) Plots of the filling fraction (a) and the size parameter (b), required to achieve 
IA in a core shell structure immersed in a $N_b=1.5$ background material. The core is considered to be a
lossless high index material, $N_c=2.8$, and the shell either silver (red-solid line) or gold 
(blue-dashed line).}
\end{figure}

We apply the graphic solution method described in the previous paragraph to the core-shell IA design consisting of a
silver or gold metallic layer around a lossless high index core of $N_c=2.8$ in a background medium of index
$N_b=1.5$. The results are illustrated in Fig.(\ref{fig:core}). The filling fractions of the shell material and the radius 
of the particle are given as a function of the vacuum wavelength of the incident field. We remark
that the chosen core-shell design with a high index center and a metallic shell allows one to design IA for any
wavelength higher than the homogeneous sphere solutions given in Table (\ref{tab:glassback}). It was also possible to 
design IA for other geometries, like a dielectric shells around a metallic core, but the metallic shell 
design illustrated here tended to give the smallest IA solutions and seems more realizable from a practical standpoint.

\section{Conclusions}
We saw in this work that the IA criteria takes the form of a limit behavior of an absorption cross section channel. 
Precise electromagnetic calculations in the complex frequency plane were shown to be able to predict this
phenomenon, and new analytic expressions were showed to accurately predict the lowest order
electric and magnetic modes. Precise calculations also indicated the intriguing possibility of observing IA in high 
index absorbing materials and not just in metals.

We showed that IA should be observable at visible frequencies for homogeneous spheres composed of gold
or silver plasmonic particles at given frequencies and particle sizes. The signature of IA in terms of cross sections
was also presented and should prove useful in experiments. The methods and formulas developed in this work led to a
simple scheme for approximately predicting core-shell geometries supporting effective IA conditions throughout the
visible spectrum. Experimental observations of IA for such sub-wavelength particles, could help to better understand
IA and its experimental signature in quantum systems.

%\begin{acknowledgement}
The authors would like to thank Guillaume Baffou and Xavier Zambrana for interesting discussions and 
helpful remarks.
The research leading to these results has received funding from the European Research Council under the European Union's Seventh Framework Programme (FP7/2007-2013) / ERC Grant agreements 278242 (ExtendFRET), from the French Agence Nationale de la Recherche under Contract No. ANR-11-BS10-002-02 TWINS, and from the A*MIDEX project (n$^{\circ}$ ANR-11-IDEX-0001-02) funded by the Investissements d'Avenir French Government program managed by the French National Research Agency (ANR).
%\end{acknowledgement}

%\bibliographystyle{plain}
%\bibliographystyle{revtex}
%\bibliographystyle{unsrt}
%\bibliography{BrBib}

\end{document}